\documentclass{elsart}
\journal{Physics Letters B}

\usepackage{epsfig}
\usepackage{rotate}
\usepackage{color}

\usepackage{amssymb}
\usepackage{amsmath}
\usepackage{amscd}

\def\Journal#1#2#3#4{{#1} {\bf #2}, (#4) #3}

\def\NPB{{\em Nucl. Phys.} B}
\def\PLB{{\em Phys. Lett.}  B}
\def\PRL{\em Phys. Rev. Lett.}
\def\PRD{{\em Phys. Rev.} D}
\def\ZPC{{\em Z. Phys.} C}
\def\EPJ{{\em Eur. Phys. J.} C}
\def\JPG{\em Journ. Phys. G.}

\def\PR{\em Phys. Rep.}
\def\NPBps{{\em Nucl. Phys.} B {\em Proc. Suppl.}}

\newcommand{\bmath}{\begin{displaymath}}
\newcommand{\emath}{\end{displaymath}}
\newcommand{\be}{\begin{equation}}
\newcommand{\ee}{\end{equation}}

\newcommand{\kl}{$\rm K_{L} $}

\newcommand{\ke}{$\rm K_{e3}$}

\newcommand{\kppp}{$\rm K_{3\pi}$}
\newcommand{\kppplong}{$\rm K_{L}\rightarrow~\pi^{+}~\pi^{-}~\pi^{0}$}
\newcommand{\kpp}{$\rm K_{2\pi}$}
\newcommand{\kpplong}{$\rm K_{L}\rightarrow~\pi^{+}~\pi^{-}$}
\newcommand{\kmzero}{$\rm K^{0}_{\mu3}$}
\newcommand{\km}{$\rm K_{\mu3}$}
\newcommand{\kmlong}{$\rm K_{L}\rightarrow~\pi^{\pm}~\mu^{\mp}~\nu_{\mu}$}
\newcommand{\kell}{$\rm K_{\ell 3} $}
\newcommand{\vus}{$\vert V_{\it us} \vert$}
\newcommand{\lp}{$\lambda_{+}$}
\newcommand{\lz}{$\lambda_{0}$}

\definecolor{darkred}{rgb}{0.5,0.0,0.0}

\begin{document}
\begin{frontmatter}
                                                                                

\title{\boldmath {Measurement of \kmzero~ form factors }} 
\collab{NA48 Collaboration}

\author{A.~Lai},
\author{D.~Marras}
\address{Dipartimento di Fisica dell'Universit\`a e Sezione dell'INFN di Cagliari, \\ 
I-09100 Cagliari, Italy} 
\author{A.~Bevan},
\author{R.S.~Dosanjh\thanksref{threfRAL1}},
\author{T.J.~Gershon\thanksref{threfRAL2}},
\author{B.~Hay},
\author{G.E.~Kalmus},
\author{C.~Lazzeroni},
\author{D.J.~Munday},
\author{E.~Olaiya\thanksref{threfRAL}},
\author{M.A.~Parker},
\author{T.O.~White},
\author{S.A.~Wotton}
\address{Cavendish Laboratory, University of Cambridge, Cambridge, CB3~0HE, U.K.\thanksref{thref3}}
\thanks[threfRAL1]{Present address: Ottawa-Carleton Institute for Physics, Carleton University, Ottawa, Ontario K1S 5B6, Canada}
\thanks[threfRAL2]{Present address: High Energy Accelerator Research Organization (KEK), Tsukuba, Japan}
\thanks[threfRAL]{Present address: Rutherford Appleton Laboratory, Chilton, Didcot, Oxon, OX11~0QX, U.K.}
\thanks[thref3]{Funded by the U.K.\ Particle Physics and Astronomy Research Council}
% ...


\begin{abstract}
\noindent This paper reports on a new high precision measurement 
of the form factors of the  \kmlong~ decay.
The data sample of about 2.3$\times 10^{6}$ events was
recorded in 1999 by the NA48 experiment at CERN.
\noindent Studying the Dalitz plot density we measured a linear,
$\lambda^{'}_{+} = (20.5\pm 2.2_{stat} \pm 2.4_{syst})\times 10^{-3}$, 
and a quadratic, 
$\lambda^{''}_{+} = (2.6\pm 0.9_{stat} \pm 1.0_{syst})\times 10^{-3}$
term in the power expansion of the vector form factor. No evidence
was found for a second order term for the scalar form factor;
the linear slope was determined to be
$\lambda_{0} = (9.5\pm 1.1_{stat} \pm 0.8_{syst})\times 10^{-3}$. 
\noindent Using a linear fit our results were:
$\lambda_{+} = (26.7\pm 0.6_{stat} \pm 0.8_{syst} )\times 10^{-3}$ and,
$\lambda_{0} = (11.7\pm 0.7_{stat} \pm 1.0_{syst} )\times 10^{-3}$.\\
\noindent A pole fit of the form factors yields: 
 $m_V = ( 905 \pm 9_{stat} \pm 17_{syst} )$ MeV/c$^2$ and
 $m_S = (1400 \pm 46_{stat} \pm 53_{syst} )$ MeV/c$^2$.
\end{abstract}

\end{frontmatter}

\section{Introduction}
\label{intro}
Since long ago~\cite{cgg} ~\kell~ decays 
($\rm K_{L}\rightarrow~\pi^{\pm}~\ell^{\mp}~\nu,~ \ell=e, \mu$) 
have offered the opportunity to test several features of the electroweak 
interactions such as the  V-A structure of weak currents, current algebra 
and the predictions of chiral perturbation theory. 
These decays have been the object of a renewed interest both
on the experimental and theoretical side since they provide the 
cleanest~\cite{leutwyler} way to extract the CKM matrix element \vus. 
\kell~decays give access to the product $f_{+}(0)$\vus, where $f_{+}(0)$,
the vector form factor at zero momentum transfer, has to be determined 
by theory.
The recent calculations at $\mathcal{O}(p^{6})$~\cite{bijnens-tala} 
in the framework of chiral perturbation theory show how $f_{+}(0)$~could 
be experimentally constrained from the slope and the curvature of the 
scalar form factor $f_{0}$~of the \km~decay.
In addition, the form factors are needed to calculate the phase space 
integrals which are another ingredient for the determination of \vus.
Finally, very recently it has been pointed out~\cite{stern_paper} how 
a precise measurement of the value of $f_{0}$ at the Callan--Treiman
point~\cite{CT} could provide a clean test of a small admixture of right
handed quark currents (RHCs) coupled to the standard W boson.\\
Until recently, the experimental knowledge~\cite{pdg2004} on 
\kell~form factors was mainly based on a certain number of old measurements 
dating back to the seventies.
The slopes obtained from \km~decays were less precise than those determined in 
 \ke~decays, and a large difference between the results from charged and 
neutral kaon decays was present. 
This difference was more pronounced for the slope $\lambda_{0}$ where, in addition, 
the situation was confused also by the presence of negative values. 
Very recent high precision experiments~\cite{istra-e,istra-m,ktev04,na48ke3ff,kloe} 
provided a more accurate determination of these quantities with values
smaller than the old PDG averages and agreement between $\rm K^{0}$~and
$\rm K^{\pm}$~measurements has been established. 
Furthermore, evidence for a quadratic term in the vector form factor was 
found, at the level of about 2 $\sigma$, by ISTRA+ in $\rm K_{e3}^{-}$ 
and by KLOE in \ke~ decays.  
A cleaner indication came also from KTeV, both in \ke~ and \km~ decays,
with a significance of about 3 $\sigma$.\\
This paper reports on a new high statistics measurement of  \km~form factors.
Following this introduction Sec.~\ref{theory} recalls the formalism about
the \km~decays, Sec.~\ref{exp} describes the experimental  set--up, 
Sec.~\ref{analysis} reports about the analysis, 
and Sec.~\ref{results} delineates the fitting procedure and the treatment 
of the systematic error.

\section{The \kmlong~decay}
\label{theory}

Only the vector current contributes to \km~decays. As a result the matrix 
element can be written in terms of two dimensionless form factors $f_{\pm}$~:\\
\be
{\frak M} = \frac{G}{\sqrt{2}}~ V_{us}~ \left[ f_{+}(t)
 \left(P_{K}+P_{\pi}\right)^{\mu}
 \bar u_{\ell} \gamma_{\mu} (1+\gamma_{5}) u_{\nu} +
 f_{-}(t) ~ m_{\ell} \bar u_{\ell} (1+\gamma_{5}) u_{\nu}  \right],
\ee
where $P_{K}$ and $P_{\pi}$ are the kaon and pion four momenta, respectively,
$\bar u_{\ell}$ and $u_{\nu}$ are the lepton spinors, $m_{\ell}$ is the
lepton mass and 
$t = (P_{K} - P_{\pi})^{2} = m_{K}^{2} +  m_{\pi}^{2} - 2P_{K} \cdot P_{\pi} 
= q^{2}$
is the square of the four--momentum transfer to the lepton pair.
The form factor $f_{-}(t)$ is related to a scalar 
term proportional to the lepton mass and can be measured only in  
\km~decays.\\
The determination of the form factors in this analysis is based on a study
of the Dalitz plot density which can be parametrized \cite{cgg} as:
\be
\label{dalitzparam}
  \rho(E_{\mu}^{*},E_{\pi}^{*}) =
  { dN^2 (E_{\mu}^{*},E_{\pi}^{*})  \over dE_{\mu}^{*}~ dE_{\pi}^{*} }
  \propto A f_{+}^{2}(t)  + B f_{+}(t) f_{-}(t) + C f_{-}^{2}(t),
\ee
where A, B and C are kinematical terms:
\begin{align*}
A &= m_{K} ( 2 E_{\mu}^{*} E_{\nu}^{*} - m_{K} E_{\pi}^{'}) + 
     m_{\mu}^{2} (1/4~E^{'}_{\pi}-E_{\nu}^{*}),\\
B &= m_{\mu}^{2} (E_{\nu}^{*} - 1/2~E_{\pi}^{'}),\\
C &= 1/4~m_{\mu}^{2} E_{\pi}^{'}.           
\end{align*}
$E_{\mu}^{*}, E_{\pi}^{*}$ are the muon and
pion energy in the kaon center of mass (CMS) respectively. 
For the neutrino we have 
$E_{\nu}^{*} = m_{K} - E_{\mu}^{*} - E_{\pi}^{*}$
and $E_{\pi}^{'}$ is defined as:\\
\begin{equation*}
E_{\pi}^{'} = E_{\pi}^{*Max} - E_{\pi}^{*} = 
 \frac {m_{K}^{2} + m_{\pi}^{2} - m_{\mu}^{2}} {2m_{K} } - E_{\pi}^{*}.
\end{equation*}
In an alternative parametrization one can define the form factor $f_{0}(t)$:\\
\be
f_{0}(t) = f_{+}(t) + {t \over (m_{K}^{2} -  m_{\pi}^{2}) } ~f_{-}(t).
\ee
This parametrization is preferred because $f_{+}$ and $f_{ 0}$ are related 
to the vector ($1^{-}$) and scalar ($0^{+}$) exchange to the lepton system, 
respectively and are less correlated than in the previous case.
The expansion in powers of $t$ of the form factors is often stopped at
the linear term:\\
\be
\label{linear-exp}
f_{\pm, 0}(t) =  f_{\pm, 0}(0) ~
   \left( \rule{0mm}{5mm} 1 + \lambda_{\pm, 0}~ t/m_{\pi}^{2} \right).
\ee
The assumption that $f_{+}$ and $f_{ 0}$ are linear in $t$ implies 
that $f_{+}(0) = f_{0}(0) $ so that $f_{-}$ does not diverge at $t=0$.\\
The form (\ref{linear-exp}) is usually adopted as consequence of the smallness
of data samples from the past experiments rather than on physical 
motivations. Nowadays, with higher statistics, it is becoming customary
to search for a second order term in the form factors expansion:\\
\be
f_{+,0}(t) =  f_{+}(0) ~
   \left[ \rule{0mm}{5mm} 1 + \lambda^{'}_{+,0}~ t/m_{\pi}^{2} + 
   \frac{1}{2}~ \lambda^{''}_{+,0}~ (t/m_{\pi}^{2})^{2} \right].
\ee
The weak interaction of hadron systems at low energies can also be
described in terms of couplings of mesons to the weak gauge bosons
(pole model, meson dominance~\cite{MD}). 
In this framework the form factors acquire a physical meaning
since they can be related to the exchange of the lightest $\rm K^{*}$
resonances which have spin--parity $1^{-}$/$0^{+}$ and mass
$m_{V}$/$m_{S}$, respectively:\\
\be
f_{+}(t) =  f_{+}(0) ~\frac{m^{2}_{V}}{m^{2}_{V}-t};~~~~~~~~~~~~~~
f_{0}(t) =  f_{+}(0) ~\frac{m^{2}_{S}}{m^{2}_{S}-t}.
\ee
Recently new parametrizations of the vector~\cite{stern_private} and 
scalar~\cite{stern_paper} form factors based on dispersion relations 
subtracted twice have been proposed:\\
\be
f_{+}(t) = f_{+}(0)\exp\Bigl{[}\frac{t}{m^{2}_{\pi}}(\Lambda_{+} + H(t))\Bigr{]};~~~~~
f_{0}(t) = f_{+}(0)\exp\Bigl{[}\frac{t}{(m_{K}^{2} -  m_{\pi}^{2})}(\ln C - G(t))\Bigr{]},
\ee
here $\Lambda_{+}$ is a slope parameter and 
$\ln C = \ln[ f_{0}(m_{K}^{2} -  m_{\pi}^{2}) ]$ is the logarithm of
the value of the scalar form factor at the Callan--Treiman point. 
For the dispersive integrals $H(t)$ and $G(t)$ accurate polynomial approximations 
have been derived.

\section{Experimental set--up}
\label{exp}
For the measurement reported here the data were taken during a dedicated 
run period in September 1999.
A pure $\rm K_{L}$ beam was produced by 450 GeV/c protons from the 
CERN SPS hitting a beryllium target. The decay region was contained in
a 90 m long  evacuated tube and was located 126 m downstream the 
target after the last of three collimators.\\
The NA48 detector was originally designed for a precise measurement of 
direct CP violation in the neutral kaon decays to two pions. 
A detailed description can be found in \cite{na48det}; only the 
main components relevant for this measurement are described here:\\
{\bf Magnetic Spectrometer.~} It was contained in a helium tank kept
at atmospheric pressure and consisted of four drift chambers and a magnet.
Each chamber had four views (x, y, u, v) each of which had two planes of 
sense wires. The spatial resolution per projection was 100 $\mu$m and the 
time resolution was 0.7 ns. The magnet, placed between the second and the 
third chamber, was a dipole with a transverse momentum kick of 265 MeV/c.
The momentum resolution of the spectrometer was ($p$ in GeV/c):\\
\begin{equation*}
 \frac {\sigma_{p}} {p} = 0.48 \% \oplus 0.009~p \%. 
\end{equation*}
{\bf Hodoscope.~} Located downstream of the spectrometer, it was 
used to provide a precise time reference for tracks. It consisted 
of two orthogonal planes of scintillators segmented in horizontal
and vertical strips and arranged in four quadrants. The time resolution 
per track was about 200 ps. The coincidence of signals from quadrants was
used in the first level trigger for events with charged particles.\\
{\bf Electromagnetic calorimeter.~} This was a quasi--homogeneous
liquid krypton calorimeter (LKr) with projective tower structure made 
by Be--Cu 40 $\mu$m thick ribbons extending from the front to the back of 
the device in  a small accordion geometry. 
The 13248 read--out cells had a cross-section of $2 \times 2$ cm$^2$. 
The energy resolution was parametrized as ($E$ in GeV):\\
\begin{equation*}
 \frac {\sigma_{E}} {E} = \frac {(3.2\pm0.2)\%}{\sqrt{E}} \oplus
             \frac {(9\pm1)\%}{E} \oplus (0.42\pm0.05)\%.  
\end{equation*}
{\bf Muon system.~} The muon system (MUV) was located between the hadron 
calorimeter and the beam dump and consisted of three planes of 
scintillators each shielded by a 80 cm thick iron wall. The first two 
planes were made of 25 cm wide horizontal and vertical scintillators 
strips. The strips overlapped slightly in order to ensure  full coverage 
over the whole area of $2.7 \times 2.7 $ m$^{2}$. The third plane had 
horizontal strips 44.6 cm wide. The central strip was split with a gap 
of 21 cm to  accomodate the beam pipe. All counters (apart the central 
ones) were read out at  both sides. The inefficiency of the system was at 
the level of one per mill  and the time resolution was below 1 ns. 
The passage of particles in the MUV produces ''hits'', i.e. a coincidence
between an horizontal and a vertical counter which defines a 
$25 \times 25$ cm$^2$  region.\\
{\bf Trigger.~} The acquisition of events was driven by a two level
trigger. In the first level the presence of at least two hits in the 
hodoscope was requested. In the second level trigger the spectrometer
was used to reconstruct tracks and a vertex made of opposite charge tracks 
in the decay region was required. To measure the trigger efficiency, a
control trigger was implemented using the first level trigger 
properly downscaled.

\section{Data analysis}
\label{analysis}
\subsection{Event selection}
\label{selection}
The data sample consists of about $10^{8}$ triggers recorded alternating 
the polarities of the magnetic field of the spectrometer. 
To identify the \km~decays the following selection criteria were applied 
to the reconstructed data.\\
The events were required to have exactly two tracks of opposite charge
forming a vertex in the decay region, defined to be between 7.5 m and 
33.5 m from the exit of the final collimator and within 2.5 cm from the 
beam line. The distance of closest approach of these two tracks had to 
be less than 2 cm.\\
The difference in the track times reconstructed by the spectrometer 
had to be less than 6 ns while for the times determined by the hodoscope
a maximum difference of 2 ns was admitted.\\ Both tracks were required to be
inside the detector acceptance by demanding that their projection had 
to be inside the fiducial area of the various subdetectors.
Tracks were accepted in the momentum range between 10 and 170 GeV/c. \\
In order to allow a clear separation of showers, a minimum distance of 35 cm
between the extrapolated impact points of the tracks at the entrance of the 
LKr calorimeter was required. Furthermore to avoid problems due to the 
misreconstruction of the shower energy a minimum distance of 2 cm from 
the track impact point to a dead calorimeter cell was imposed.\\
In order to reduce the background from \kppplong~(\kppp) decays the cut 
$P_{0}^{'2} < -0.004$ (GeV/c)$^2$ was applied. 
The variable $P_{0}^{'2}$, which is computed assuming that 
the decay is a \kppp, is defined as:\\
\begin{equation*}
P_{0}^{'2} = \frac{1}{4(p_{\perp}^{2} + m_{+-}^{2} ) }
\left[ \left( m_{K}^{2} - m_{+-}^{2} - m_{\pi^{0}}^{2} \right) ^{2}
- 4 \left( m_{+-}^{2}m_{\pi^{0}}^{2} + m_{K}^{2}p_{\perp}^{2} \right) \right].
\end{equation*}
In the above formula $p_{\perp}~(m_{+-})$ is the transverse momentum 
(invariant mass) of the assumed $\pi^{+} \pi^{-}$ system relative to the
direction of the \kl. $P_{0}^{'2}$ represents the \kl~momentum in a reference 
frame in which the longitudinal component of the pion system is zero. It is
positive for \kppp~decays and negative for \kell~decays.\\
The \ke~background was suppressed using the ratio $E/p$, where $E$ is
the energy of the cluster, reconstructed in the electromagnetic calorimeter 
and associated to a track, and $p$ is the track momentum as measured by the 
spectrometer.
For both tracks $E/p$ had to be less than 0.9. The probability for
a $\pi$ to be rejected by this cut was measured to be about 1\%. \\
A track was identified as a muon when its extrapolated impact point
at the MUV could be spatially associated to a MUV hit. The distance of
association was dependent on the momentum of the track to account for multiple
scattering and measurement errors.
For this analysis in addition, other constraints were imposed: the distance 
between the track  extrapolation and the hit had to be less than 30 cm; the 
difference between the event time (determined by the charged hodoscope) and 
the muon time (determined by the MUV) had to be less than 3 ns, and finally 
also a coincidence in the MUV plane 3 was required. 
Monte--Carlo (MC) studies indicate that under these conditions
the probability to misidentify a $\mu$ for a $\pi$ is at the $10^{-5}$ level.\\
A well--known problem with $\rm K_{\ell 3}$ decays is the
quadratic ambiguity in the determination of the \kl~energy.
The $\nu$ escapes undetected in this decay and while 
the transverse component of the momentum 
($p_{\nu T}$) to the \kl~direction of flight (obtained joining 
the event vertex to the target position) is determined by the $\mu$ and $\pi$
transverse momenta, the longitudinal component ($p^{*}_{\nu L}$) can be 
determined only up to a sign representing the two possible orientations of 
the $\nu$ in the kaon CMS. 
This ambiguity leads to two solutions for kaon energy, called "low" 
($E_{L}$) and "high"($E_{H}$).  
As an additional selection criteria we required both kaon energy solutions 
to be greater than 70 GeV.\\
Finally a cut was applied on the variable $p_{\nu}^{*} - p_{\nu T}$,
where $p_{\nu}^{*}$ is the total neutrino momentum in the kaon CMS.
This quantity, clearly positive for good \km~events, is highly
sensitive to resolution effects which give rise to a moderate negative 
tail. We set a cut at $p_{\nu}^{*} - p_{\nu T}  > 7$ MeV/c, selecting
a region where the MC simulation accurately reproduces the data behaviour.
After the applications of all cuts, 2344382 \km~events were reconstructed 
from the data sample.
\begin{figure}[ht]
\begin{center}
   \epsfig{file=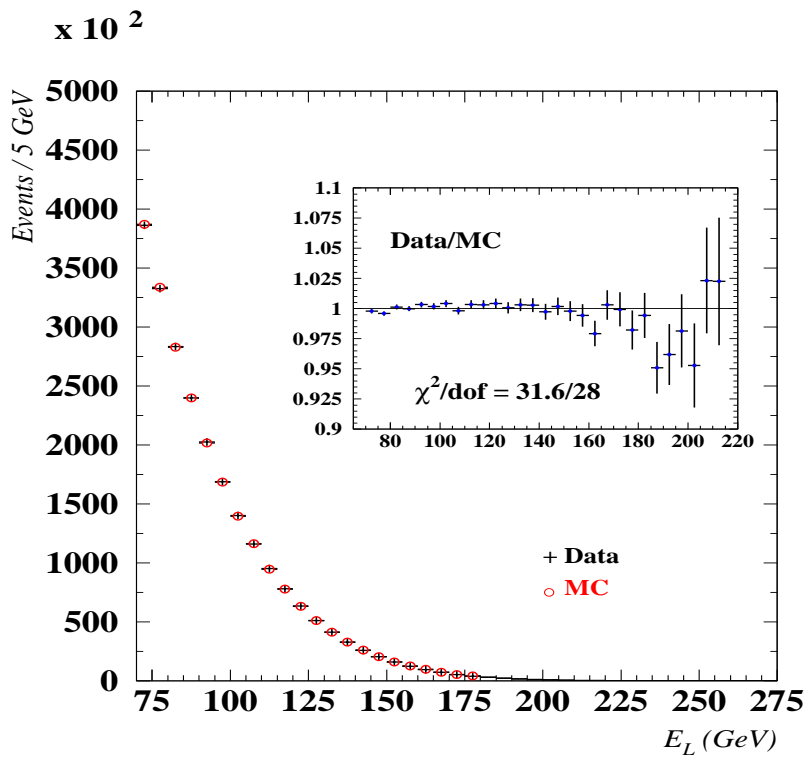,height=100mm,width=120mm}
\end{center}
\vskip -0.25cm
\caption{{\it
Comparison data--MC for the low kaon energy solution, 
the inset shows the ratio data/MC.}}
\label{pl}
\end{figure}
\subsection{Monte--Carlo simulation}
\label{MC}
\begin{figure}[htb]
\begin{center}
  \epsfig{file=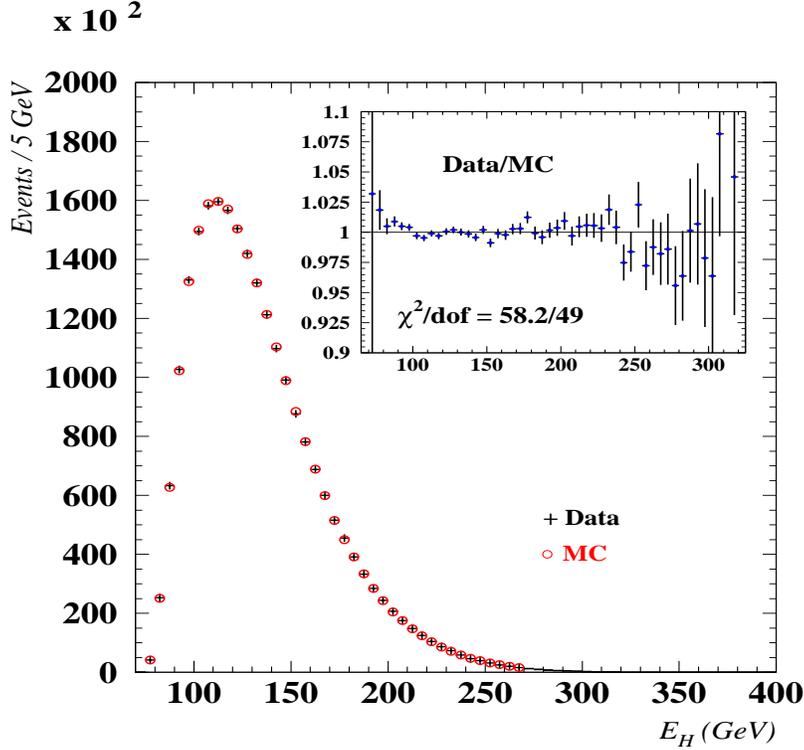,height=100mm,width=120mm}
\end{center}
\vskip -0.25cm
\caption{{\it
Comparison data--MC for the high kaon energy solution, 
the inset shows the ratio data/MC.}}
\label{ph}
\end{figure}

\begin{figure}[htb]
\begin{center}
   \epsfig{file=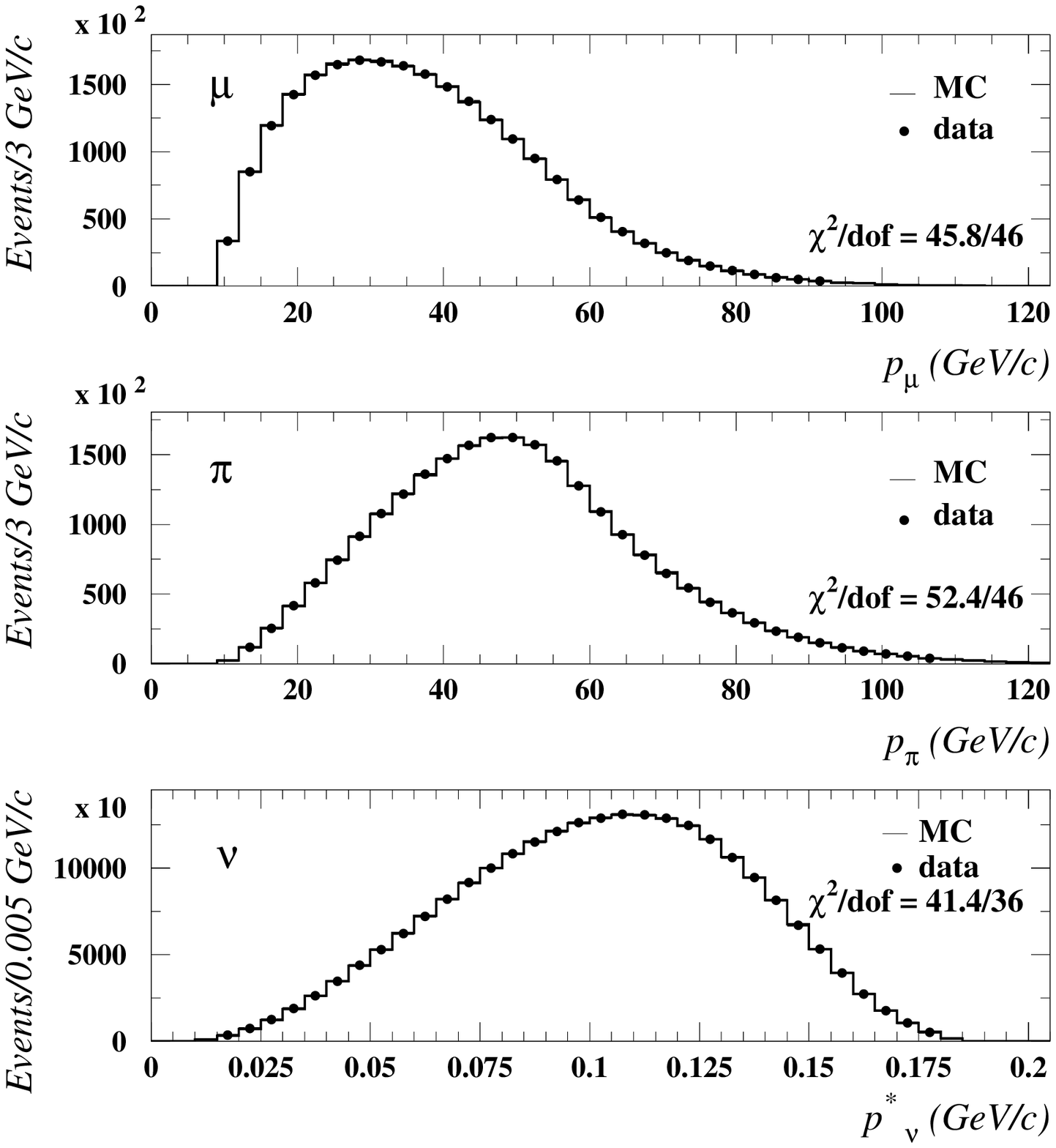,width=150mm}
\end{center}
\vskip -0.5cm
\caption{{\it
Comparison data--MC for the momentum distributions of muons, pions 
(lab system) and neutrinos (kaon CMS).}}
\label{momentum}
\end{figure}

The detector response has been simulated in details using a  
MC program based on {\tt GEANT}~\cite{geant}. Particle interactions in 
the detector  material as well as response functions of the different 
detector elements  have been taken into account in the simulation. 
Pre--generated shower libraries for photons, electrons and charged pions 
are used to describe the response of the  calorimeters. \\
To determine the detector acceptance as well as the distortion 
and losses of events on the Dalitz plot induced by the radiative effects,
the \km~decay has been simulated both at the Born level and
at the next--to--leading order (NLO). 
The acceptance suffers only from a residual dependence on the values 
(and on the type of parametrization) of 
the form factors used for the generation of the MC samples. 
To avoid any biases the samples were produced, after an iterative procedure, 
with values close to the results reported here. 
A linear parametrization was used with $\lambda_{+}^{gen}=0.0260$ and
$\lambda_{0}^{gen}=0.0120$.\\
The NLO sample was obtained using as event generator {\tt KLOR}~\cite{klor}, 
a program which numerically evaluates the radiative corrections and generates 
MC events. The simulated events underwent the same reconstruction procedure 
as the data events and the same selection cuts described in 
Sec~\ref{selection} were applied. 
These two MC samples provide a statistics which is one  order of magnitude 
larger than the data one. \\
A third, smaller, \km~sample was generated with full simulation of the 
showers in the calorimeters and was used to model the multiple scattering 
in the MUV. For detailed studies of the background samples of \ke, 
\kppp~and \kpplong~(\kpp) decays decays were  produced.\\
The \kl~energy spectrum was extracted from the data by using the data
distributions of the low and high energy solutions
and the probability matrix, obtained from MC, which relates
$E_{L}$ and $E_{H}$ to the true kaon energy.\\
To show the quality of the MC simulation the comparison of data and MC 
for some relevant kinematical quantities are shown in Fig.~\ref{pl} to 
\ref{momentum}.
\subsection{Backgrounds}
\label{backgrounds}
The \ke, \kppp~and \kpp~decays are the major sources of background.
A \ke~event can fake a genuine \km~when the $\pi$~decays into a muon 
and the  $e$ has the $E/p$ requested for the identification of a 
$\pi~(E/p < 0.9)$.
To determine this source of contamination in the selected \km~sample, we 
used the $E/p$ distributions of tracks which pass all cuts, but not 
considering $E/p$. 
The electron signal is obtained by fitting this distribution
around  the value of 1. The integration of the fitted function into the 
''pion'' region allows to determine a value for the \ke~contamination of:
\be
\mathcal{P}^{cont}_{Ke3} = (6.59 \pm 0.09) \times 10^{-4}.
\label{ke3cont}
\ee
The \kppp~decays (followed by the decay of one of the two charged
$\pi$) are strongly suppressed by the $P_{0}^{'2}$ cut. 
To determine the residual contamination the selected \km~events
undergo a \kppp~selection procedure: in the presence of
clusters in the LKr not associated to the tracks, an attempt to reconstruct 
a $\pi^{0}$ is made. In case the two photons reconstruct the $\pi^{0}$
mass within a window of $\pm6$ MeV/c$^2$, the invariant mass of the 
two tracks (assumed to be pions) and  the $\pi^{0}$ is evaluated and 
if it falls in an interval of $\pm9$ MeV/c$^2$ around the  $K^{0}$ mass 
the event is assumed to be a \kppp~decay.
The number of these \kppp~background events, corrected for their
acceptance, allows to estimate for the \kppp~contamination the value:
\be
\mathcal{P}^{cont}_{3\pi} = (6.31 \pm 0.16) \times 10^{-4}.
\label{k3picont}
\ee
Another source of background stems from the \kpp~decay with subsequent 
$\pi$ decay in flight or pion punch--through in the iron of the MUV. 
Using the \kpp~MC sample this contamination is estimated to be:
\be
\mathcal{P}^{cont}_{2\pi} = (5.63 \pm 0.16) \times 10^{-4}. 
\ee
This background source turns out to be the most dangerous one since the 
\kpp~events populate a narrow region (the top right corner) of the \km~Dalitz plot
introducing appreciable distortions. The \kppp~events instead populate the bottom 
left region of the plot; being not much concentrated, they induce a smaller effect.
Finally the \ke~events are distributed randomly on the plot and their effect 
is negligible.\\
Background events from \kpp~and \kppp~will be subtracted from the data
while the effects of \ke~events will be included in the treatment of the 
systematic uncertainty related to the background (Sec.\ref{systematics}).

\section{Fitting procedure and results}
\label{results}
\subsection{Fitting procedure}
\begin{figure}[hbt]
  \hskip -1.5cm
  \begin{tabular}{cc}
   \mbox{\epsfig{file=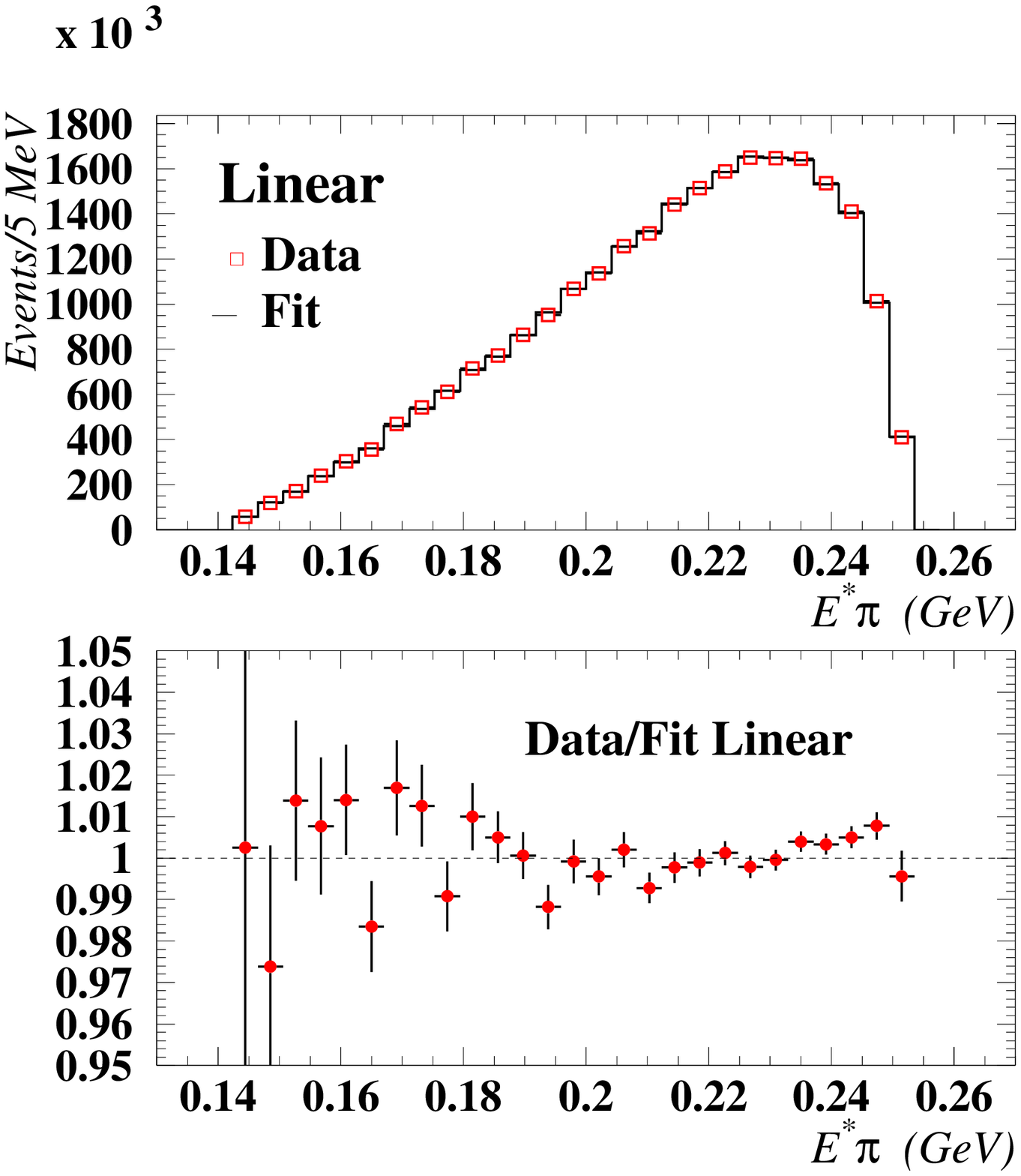,height=90mm,width=90mm}} &
   \mbox{\hskip -1.5cm \epsfig{file=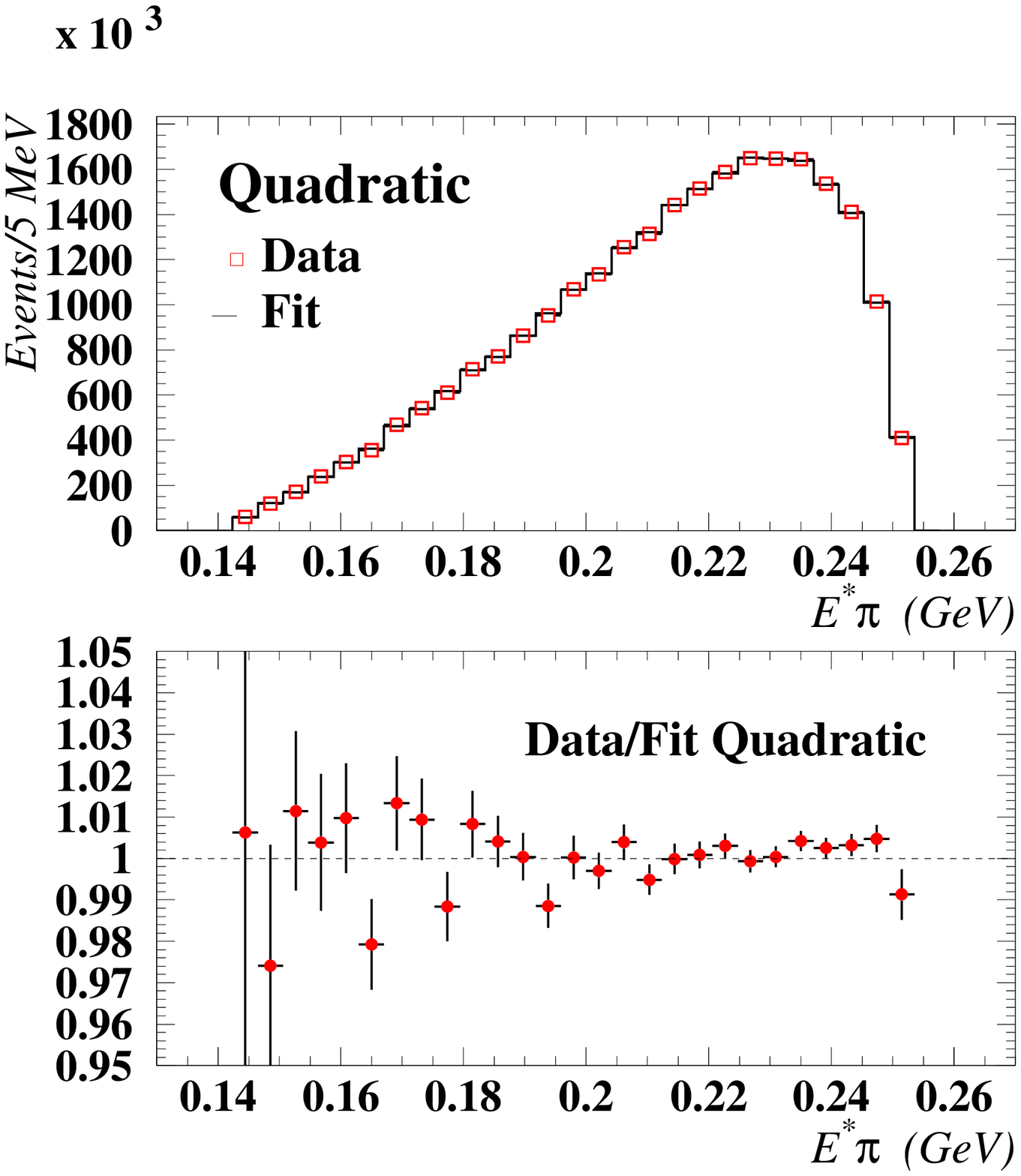,height=90mm,width=90mm}}\\
   \mbox{\epsfig{file=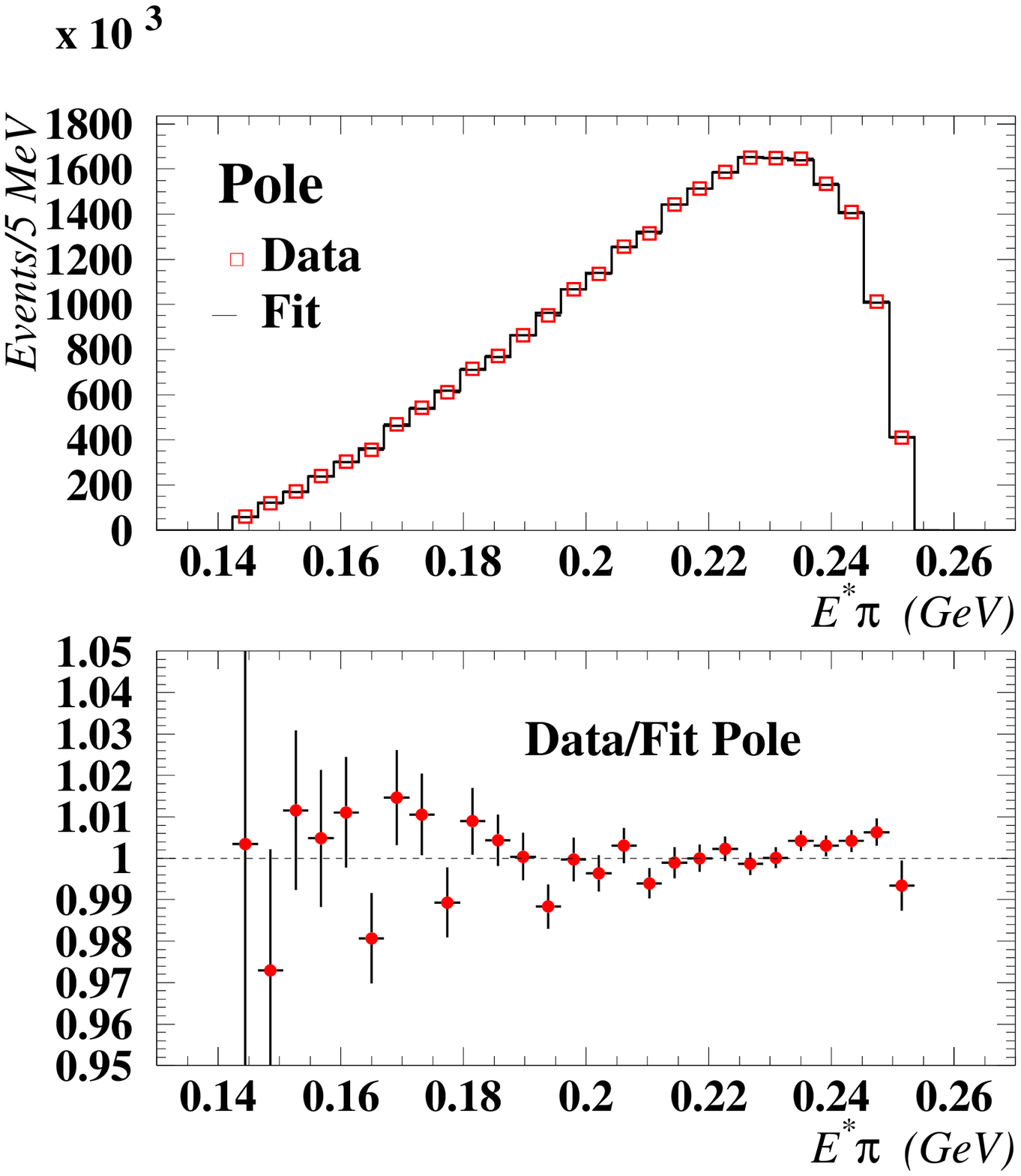,height=90mm,width=90mm}} &
   \mbox{\hskip -1.5cm \epsfig{file=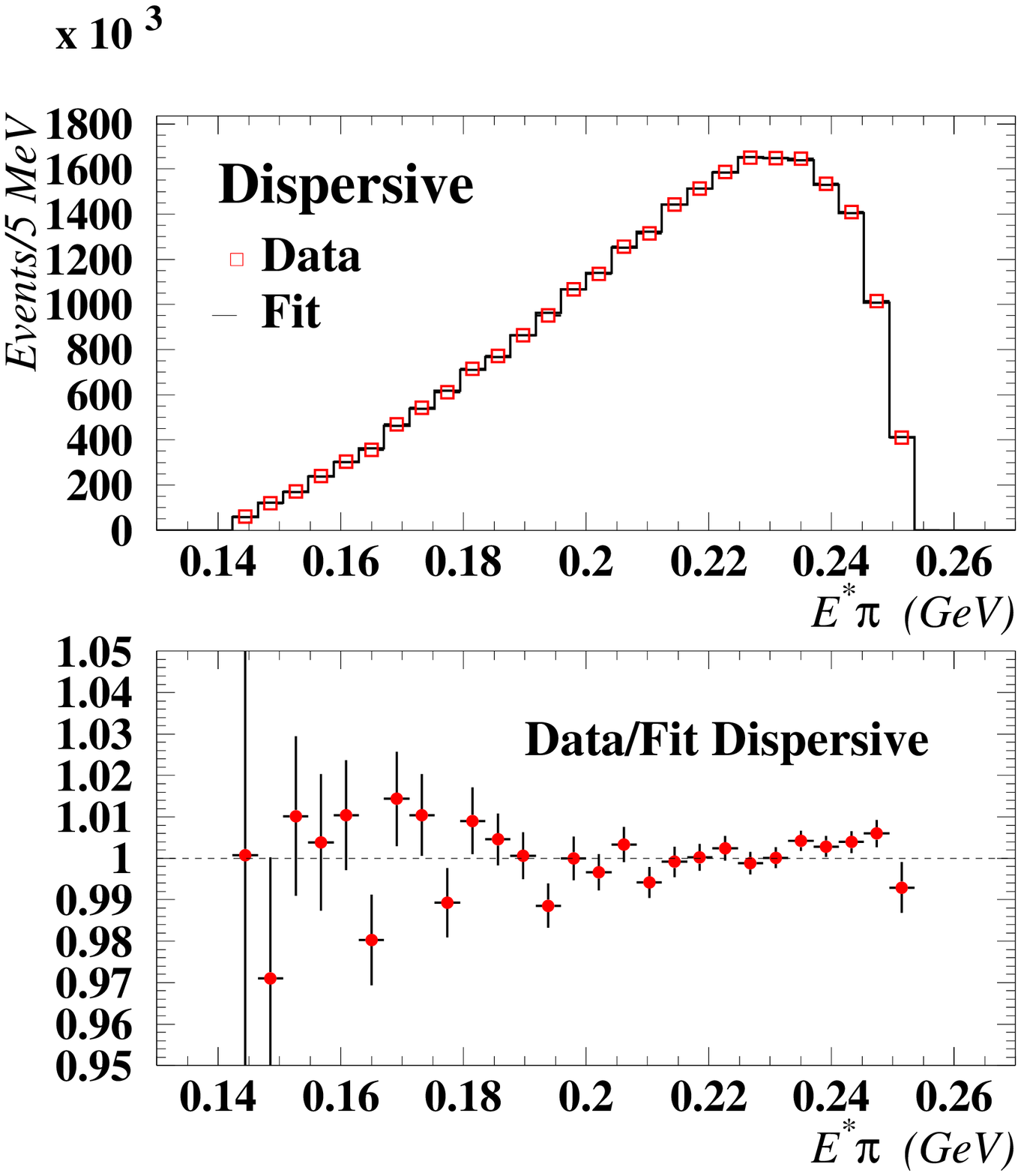,height=90mm,width=90mm}}\\
  \end{tabular}
  \caption{\it Comparison Data--Fits and Data/Fit ratios for the four parametrization 
  used in the analysis. For visualization purposes the various Dalitz plots
  distributions have been projected onto the $E_{\pi}^{*}$ axis.} \label{datafit}
 \end{figure}

The measurement reported here is based on the study of the Dalitz
plot density. 
\begin{figure}[htb]
\begin{center}
   \epsfig{file=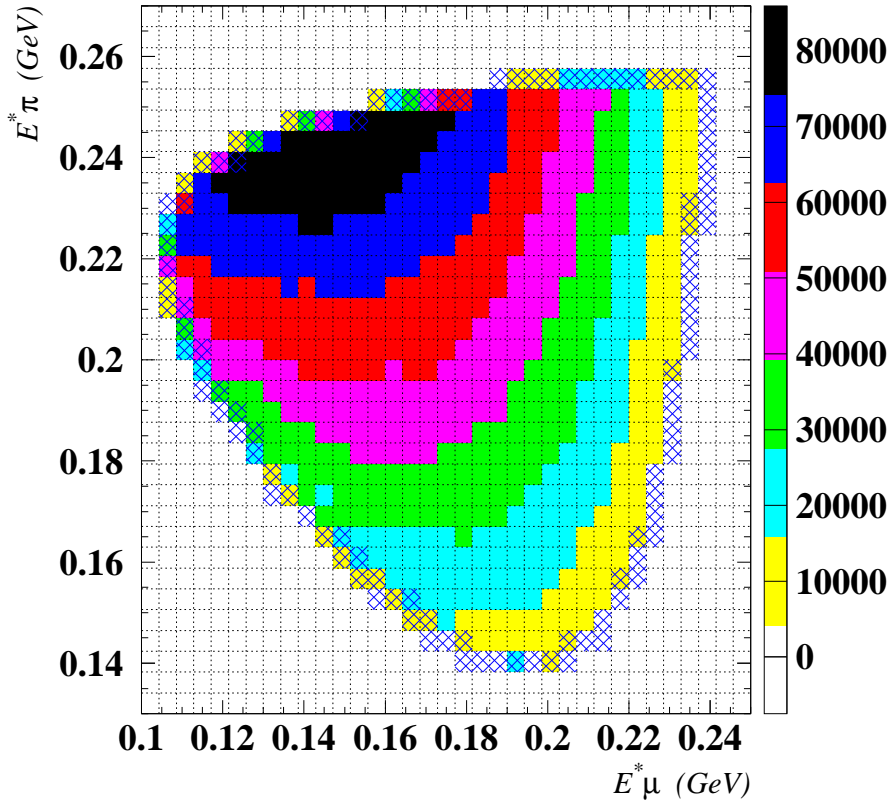,height=120mm,width=120mm}
\end{center}
\caption{{\it
Dalitz plot distribution of the \km~events corrected for radiative 
effects and acceptance; the shaded cells are not used for the fit.}}
\label{dplot_fig}
\end{figure}
As mentioned before, the ambiguity in the determination of
the kaon energy leads to two solutions for the \kl~energy and for 
the CMS energies of the $\mu$ and the $\pi$. Consequently each event 
has a double location on the Dalitz plot.
We chose to evaluate $E_{\mu}^{*}$ and $E_{\pi}^{*}$ by using only the 
low kaon energy solution. According to the MC simulation,
this corresponds to the most probable solution,
being in 61\% of cases the correct one. The Dalitz plot was divided
into cells with a dimension of about $4 \times 4$ MeV$^2$ 
(see Fig.~\ref{dplot_fig}). About 39\% of the events are reconstructed 
exactly in the same cell where they were generated, while this figure
drops to 22\%  if the high solution is used. 
To extract the form factors we fit the data Dalitz plot, corrected
for acceptance and radiative effects, to the Born level prediction.
The acceptance, in the  $i$--th cell of the plot, $\epsilon_{i}$, 
is defined as the ratio of the number of reconstructed events
(evaluated using the low energy solution)  to the number of 
generated events (evaluated using the true kaon energy) in that cell.
We note that this definition of acceptance accounts also for the migration
of events induced by the use of the low solution only.\\
The correction (for the  $i$--th cell of the plot) due to the radiative
effects is $(1 + \delta)_{i}$ and is evaluated by taking the ratio 
between the number of reconstructed events from the MC--NLO sample 
and the number of reconstructed events from the MC--Born one.\\
The number of events, corrected for acceptance and radiative effects, 
in a given cell of the plot is therefore:
\be
 N_{i} =
 \frac{ N_{i}^{Rec} }{ \rule{0mm}{3mm} \epsilon_{i} (1 + \delta)_{i} },
\ee
where $ N_{i}^{Rec}$ is the number of reconstructed and background 
subtracted data events.\\
The form factors were determined by fitting with the {\tt MINUIT}~\cite{minuit}
package the Dalitz plot distribution, corrected for acceptance and 
radiative effects, to the parametrization reported in Eq.~\ref{dalitzparam}.
The cells crossed by the Dalitz plot boundary are excluded from the fit
(see Fig.~\ref{dplot_fig}).
Various $t$ dependences of the form factors were considered: linear, 
quadratic, pole and dispersive. The fit results are listed in 
Table~\ref{table:fit_results}; the correlations among the fitted form factors
parameters are shown in Table~\ref{table:correl}. The comparison Data--Fit 
are shown in Fig.~\ref{datafit}.\\
\begin{table}[h]
\begin{tabular}{lccccc}
 \hline
Linear ($\times 10^{-3}$) &  \lp & \lz  &  & $\chi^2$/ndf \\
& 26.7$\pm$0.6 & 11.7$\pm$0.7 & &  604.0/582 \\

 \hline
Quadratic ($\times 10^{-3}$) & $\lambda^{'}_{+}$  & $\lambda^{''}_{+}$ & 
\lz  & $\chi^2$/ndf \\
& 20.5$\pm$2.2 & 2.6$\pm$0.9 & 9.5$\pm$1.1 &  595.9/581 \\

 \hline
Pole (MeV/c$^2$) & $m_V$  & $m_S$  & & $\chi^2$/ndf \\ 
& 905$\pm$9 & 1400$\pm$46 &  & 596.7/582\\

 \hline
Dispersive ($\times 10^{-3}$) & $\Lambda_{+}$  & $\ln C$ &  & $\chi^2$/ndf \\
 &  23.3$\pm$0.5  & 143.8$\pm$8.0 & &  595.0/582 \\

 \hline
\end{tabular}
\medskip
\caption{{\it Form factors fit results for linear, quadratic pole and dispersive 
parametrizations. The quoted errors are the statistical ones.}}
\label{table:fit_results}
\end{table}
\noindent We also fitted for a possible quadratic term in the scalar form 
factor and found 
$\lambda^{''}_{0} = (1.1\pm1.3)\times 10^{-3}$
indicating that the linear assumption is sufficient to describe 
this form factor.
\begin{table}[htb]
\begin{tabular}{lccc}
 \hline
Linear &     &  \lz   & \\
       & \lp & -0.40 & \\
 \hline
Quadratic &  &   $\lambda^{''}_{+}$   &    \lz  \\
       & $\lambda^{'}_{+}$ & -0.96  &  0.63 \\
       & $\lambda^{''}_{+}$ & $-$  &  -0.73 \\
 \hline
Pole   &       &  $m_S$ & \\
       & $m_V$ & -0.47 & \\
 \hline
Dispersive &     &   $\Lambda_{+}$   & \\
       &  $\ln C$ &  -0.44 & \\
\hline
\end{tabular}
\medskip
\caption{{\it Correlation coefficients among the fit parameters 
for the linear, quadratic, pole and dispersive parametrizations. }}
\label{table:correl}
\end{table}
\noindent In the expansion of the vector form factor instead, evidence is present 
for the existence of both a linear and a quadratic term.
We notice also a remarkable shift in the value of \lz~as consequence
of the presence of the quadratic term in the vector form factor expansion.
The value of $m_V$ and $m_S$, obtained with the pole fit 
are found to be consistent with the $\rm K^{*}(892)$ and
$\rm K^{*}(1430)$ masses, respectively.\\
As a cross--check we extracted the linear form factors using a
$ \chi^{2} $ built by comparing the data Dalitz plot distribution, 
corrected for radiative effects only, with a set of Born 
level plots of reconstructed MC events. Each MC Dalitz plot distribution 
was produced with different form factors values by proper re--weighting
of the events of the reference MC--Born sample. The form factors
values are extracted by minimizing the 
$ \chi^{2}(\lambda_{+},~\lambda_{0}) $ function constructed in this way.
The results obtained with this method are less accurate than those
provided by the fit procedure but fully unbiased with respect to the choice 
of the form factors values used to generate the MC sample.
These results are in perfect agreement with the ones obtained with the
fit procedure, indicating the absence of such kind of bias in the analysis procedure.\\ 
To check the fit procedure we fitted MC events, using the reference MC sample 
(generated with linear parametrization) and smaller samples generated with quadratic 
and pole parametrizations. In all the three cases the input form factors were correctly 
reproduced at the end of the process, indicating the absence of any bias in the 
fit procedure.\\

\subsection{Systematic uncertainties}
\label{systematics}
Various sources of systematic uncertainties in the determination of the 
form factors have been investigated. Their individual contributions
are reported on Table~\ref{table:syst_error} together with the effects 
related to the background contamination. The total error was obtained 
by combining the individual errors in quadrature.\\
\begin{table}[htb]
\hskip -1cm
\begin{tabular}{lccccccccc}
  \hline
 & $\Delta$\lp & $\Delta$\lz & $\Delta \lambda_{+}^{'}$ &
   $\Delta \lambda_{+}^{''}$  & $\Delta$\lz &
   $\Delta m_{V}$ & $\Delta m_{S}$ &
   $\Delta \Lambda_{+}$ & $\Delta \ln C$  \\
& \multicolumn{5}{c} {$\times 10^{-3}$} & \multicolumn{2}{c}{ MeV/c$^{2}$} 
& \multicolumn{2}{c} {$\times 10^{-3}$}\\
  \hline
  & & & & & & & & &\\
Background            & $\pm$0.0 & $\pm$0.1 & $\pm$0.2 & $\pm$0.1 & $\pm$0.0 & $\pm$0  & $\pm$5 
                      & $\pm$0.0 & $\pm$1.2 \\
Acceptance            & $\pm$0.4 & $\pm$0.5 & $\pm$0.7 & $\pm$0.4 & $\pm$0.4 & $\pm$7  & $\pm$22 
                      & $\pm$0.4 & $\pm$5.0 \\
$TRK_{dist}$ @ LKr    & $\pm$0.4 & $\pm$0.4 & $\pm$0.5 & $\pm$0.4 & $\pm$0.3 & $\pm$10 & $\pm$20 
                      & $\pm$0.4 & $\pm$5.4 \\
$P_{MIN}$             & $\pm$0.1 & $\pm$0.3 & $\pm$0.4 & $\pm$0.1 & $\pm$0.3 & $\pm$1  & $\pm$20 
                      & $\pm$0.1 & $\pm$3.1 \\
$P_{\nu}^{*} - P_{\nu T}$ 
                      & $\pm$0.2 & $\pm$0.2 & $\pm$0.5 & $\pm$0.2 & $\pm$0.2 & $\pm$6  & $\pm$10 
                      & $\pm$0.2 & $\pm$2.2 \\
\kl~ spectrum         & $\pm$0.2 & $\pm$0.4 & $\pm$0.0 & $\pm$0.0 & $\pm$0.3 & $\pm$4  & $\pm$20 
                      & $\pm$0.2 & $\pm$4.1 \\
HIGH solution         & $\pm$0.3 & $\pm$0.0 & $\pm$0.6 & $\pm$0.2 & $\pm$0.2 & $\pm$8  & $\pm$12 
                      & $\pm$0.4 & $\pm$1.9 \\
MUV reconstruction    & $\pm$0.1 & $\pm$0.1 & $\pm$0.1 & $\pm$0.0 & $\pm$0.1 & $\pm$2  & $\pm$5 
                      & $\pm$0.2 & $\pm$0.8 \\
Radiative corrections & $\pm$0.2 & $\pm$0.4 & $\pm$2.0 & $\pm$0.7 & $\pm$0.3 & $\pm$2  & $\pm$20 
                      & $\pm$0.1 & $\pm$4.3 \\
Cell Size             & $\pm$0.3 & $\pm$0.3 & $\pm$0.5 & $\pm$0.3 & $\pm$0.3 & $\pm$5  & $\pm$20 
                      & $\pm$0.2 & $\pm$4.0 \\
  & & & & & & & & & \\
  \hline
Total Systematic      & $\pm$0.8 & $\pm$1.0 & $\pm$2.4 & $\pm$1.0 & $\pm$0.8 & $\pm$17 & $\pm$53 
                      & $\pm$0.8 & $\pm$11.2 \\
  & & & & & & & & &\\
Statistical           & $\pm$0.6 & $\pm$0.7 & $\pm$2.2 & $\pm$0.9 & $\pm$1.1 & $\pm$9 & $\pm$46 
                      & $\pm$0.5 & $\pm$8.0 \\
  & & & & & & & & &\\
  \hline
Total Error           & $\pm$1.0 & $\pm$1.2 & $\pm$3.3 & $\pm$1.3 & $\pm$1.4 & $\pm$19 & $\pm$70 
                      & $\pm$0.9 & $\pm$13.8 \\
\end{tabular}
\medskip
\caption{{\it Systematic and total uncertainties for the four form factor
parametrizations analyzed. The systematic and statistical uncertainties 
have been added in quadrature to obtain the total error.}}
\label{table:syst_error}
\end{table}
\noindent Effects related to the background have been checked altering
the estimated contaminations by 15\% and accounting for the tiny effect 
related to \ke~events. The variations in the fit results were taken
as the systematic uncertainty.\\
Effects related to the acceptance and selection criteria have been 
checked by varying the selection cuts in a reasonable range. The largest 
fluctuations in the form factors were taken as systematic errors.\\
Effects related to the  \kl~ energy spectrum used in the MC simulations
were investigated by using the spectrum obtained from a clean sample
of \kpp~decays. The simulated events were re--weighted with the ratio
of the two spectra, and the differences in the form factor results were taken
as the systematic uncertainty.\\
To check effects related to the  use of the low kaon energy solution
the analysis was repeated using the high solution to determine the 
acceptance and radiative corrections. Also in this case the differences
with the reference fit results were taken as systematic error.\\
The inefficiency of the MUV during this run was measured by identifying the
$\mu$ according to its energy deposition in the electromagnetic and hadronic
(HAC) calorimeters. 
The MUV efficiency was found to vary between 0.97 for a 10 GeV/c and 1 
for a $\geq 20$ GeV/c muon with an average of $\epsilon_{MUV} = 0.9987\pm0.0001$.
To investigate possible biases, the inefficiency was artificially increased
by randomly rejecting events according to the momentum dependence of the efficiency 
and its value, without observing any significant effect.\\
Effects related to the MUV offline reconstruction were tested by relaxing the cut 
between the track extrapolation and the hit position and by accepting also events 
for which only plane 1 and 2, but not plane 3, had fired. This produces an increase
of 3.2\% in the statistics of the data sample; here again differences from 
the reference form factor values were taken as systematic error.\\
Effects related to the radiative corrections model used in the analysis
were tested by applying the corrections obtained with the Ginsberg~\cite{gins}
formalism, amending the error reported in Ref.~\cite{cirigliano} and allowing
for a $t$ dependence of the form factors. The differences with the reference
results were taken as an estimate of the systematic effect.\\
The effects related to the size of the cells in which the Dalitz plot was
divided were determind by reducing the cell size down to about 
$3 \times 3$ MeV$^2$, the largest fluctuations in the form factors 
were taken as systematic errors.\\
To estimate the possible influence of accidental particles, tracks outside the 
allowed time window for a match in the MUV were studied. No effect was found from 
this source. 

\section{Conclusions}
\label{conclusions}
\begin{figure}[ht]
\vskip -1.8cm
\begin{center}
   \epsfig{file=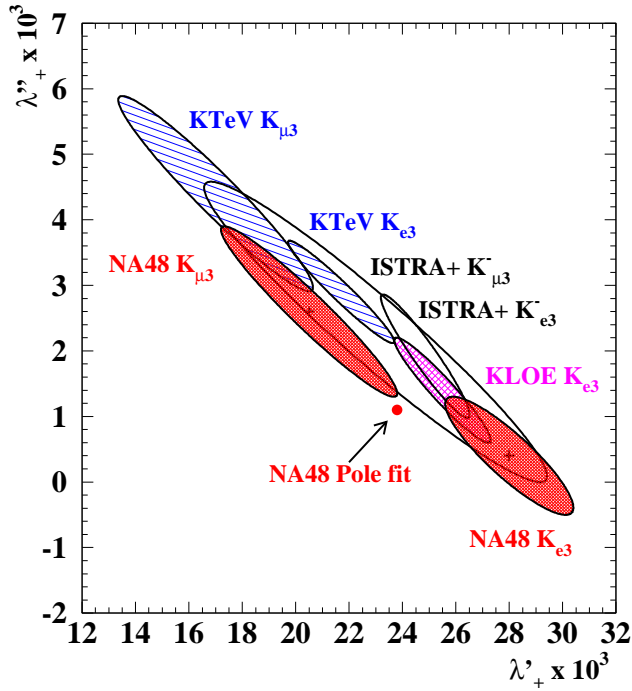,width=100mm}
\end{center}
\vskip -0.2cm
\caption{{\it
1 $\sigma$ contour plots in the plane $\lambda^{'}_{+}$--$\lambda^{''}_{+}$ showing the 
NA48 results together with those of \cite{ktev04,istra-e,kloe} for the quadratic fits of 
the \km~and~\ke~decays. 
The dot represents $\lambda^{'}_{+}$ and $\lambda^{''}_{+}$ as obtained with a Taylor
expansion of the pole parametrization.
The ISTRA+ results have been multiplied by the ratio 
$(m_{\pi^{+}}/m_{\pi^{0}})^{2}$. }}
\label{ellipse}
\end{figure}

\begin{figure}[h]
\vskip -1.25cm
\begin{center}
   \epsfig{file=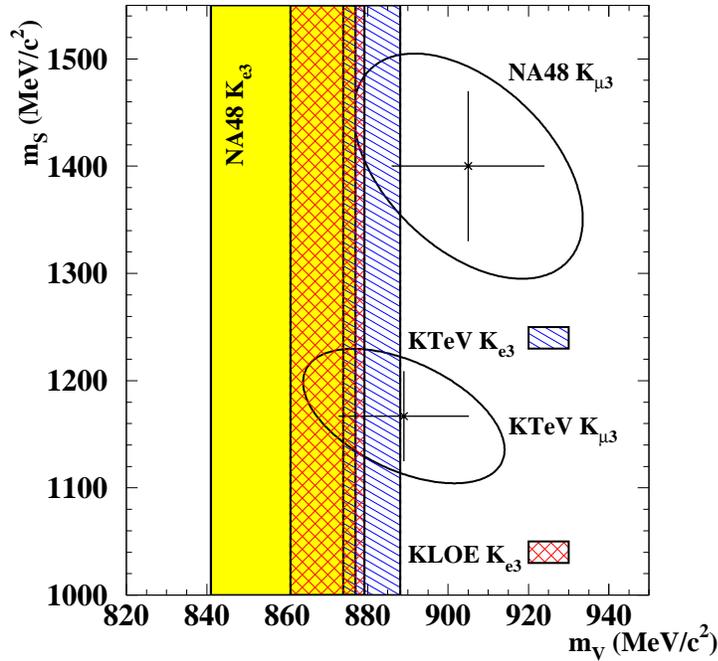,width=100mm}
\end{center}
\vskip -0.2cm
\caption{{\it
NA48 pole fits results, together with those of \cite{ktev04} and \cite{kloe},
for \km~and~\ke~decays. \ke~results appear as vertical bands on the $m_V$--$m_S$ 
plane; the ellipses shown for \km~results are the 68\% C.L. contour plots.}} 
\label{mvms}
\end{figure}

The \km~decay has been studied with the NA48 detector. 
A sample of 2.3$\times 10^{6}$ reconstructed events was analyzed
in order to extract the decay form factors. 
\noindent Studying the Dalitz plot density we measured the following
values for the form factors parameters:
$\lambda^{'}_{+}  = (20.5\pm 2.2_{stat} \pm 2.4_{syst})\times 10^{-3}$, 
$\lambda^{''}_{+} = ( 2.6\pm 0.9_{stat} \pm 1.0_{syst})\times 10^{-3}$ and
     $\lambda_{0} = ( 9.5\pm 1.1_{stat} \pm 0.8_{syst})\times 10^{-3}$.\\
Our results indicate the presence of a quadratic term in the 
expansion of the vector form factor in agreement with other
recent analyses of kaon semileptonic decays.
Fig.~\ref{ellipse} shows the comparison between the results 
of the quadratic fits as reported by the recent experiments
\cite{kloe,ktev04,na48ke3ff,istra-e,istra-m}. The 1 $\sigma$~contour plots
are shown, both for \ke~and \km~decays; the ISTRA+ results have been
multiplied by the ratio $(m_{\pi^{+}}/m_{\pi^{0}})^{2}$.
The results are higly correlated, those from this measurement and from
KTeV have a larger quadratic term and appear only in partial agreement
with the other \ke~experiments. 
We notice however that the observed spread in the 
$\lambda^{'}_{+}$, $\lambda^{''}_{+}$ figures  is greatly reduced 
if the values obtained from the Taylor expansion of the pole 
parametrization 
($\lambda^{'}_{+}  = m_{\pi}^{2}/m_{V}^{2}$;
 $\lambda^{''}_{+} = 2\lambda^{'2}_{+}$) are used.\\
Using a linear fit our results were:
$\lambda_{+} = (26.7\pm 0.6_{stat} \pm 0.8_{syst})\times 10^{-3}$ and,
$\lambda_{0} = (11.7\pm 0.7_{stat} \pm 1.0_{syst})\times 10^{-3}$.
While the result for \lp~ is well compatible with the recent (and most 
precise) KTeV measurement, the value of \lz~appears to be shifted towards 
lower values.
\noindent A pole fit of the form factors yields: 
$m_V =  (905\pm 9_{stat} \pm 17_{syst})$ MeV/c$^2$ and 
$m_S = (1400\pm 46_{stat} \pm 53_{syst}) $ MeV/c$^2$
in agreement with the $\rm K^{*}(892)$ and
$\rm K^{*}(1430)$ masses, respectively. 
In Fig.~\ref{mvms} is shown a comparison
between our results and those of \cite{ktev04} and \cite{kloe}
for this parametrization.\\
Using the recently proposed parametrization based on a 
dispersive approach, we obtain for the slope of the vector form factor:
$\Lambda_{+}= (23.3\pm 0.5_{stat} \pm 0.8_{syst})\times 10^{-3}$ 
and for the logarithm of the scalar form factor at the Callan--Treiman point:
$\ln C= (143.8\pm 8.0_{stat} \pm 11.2_{syst})\times 10^{-3}$. 
According to the model proposed in~\cite{stern_paper} the value of $\ln C$ 
can be used to test the existence of RHCs by comparing it with the Standard
Model predictions. Taking the value 
of~$\vert F_{K+} V_{us} / F_{\pi+} V_{ud} \vert $ 
from~\cite{Jamin06} and those 
of~$f_{+}(0) \vert V_{us} \vert $ and $ \vert V_{ud} \vert $
from~\cite{pdg2006} we obtain for a combination of the RHCs couplings 
and the Callan--Treiman discrepancy ($\tilde\Delta_{CT}$) the value:
$2 (\epsilon_S - \epsilon_{NS}) + \tilde\Delta_{CT} =
-0.071 \pm0.014_{NA48} \pm0.002_{theo} \pm0.005_{ext}$,
where the first error is the combination in quadrature of the
statistical and systematical uncertainties, the second one refers
to the uncertainties related to the approximations used to replace 
the dispersion integrals and the last one is due to the external
experimental input.

\newpage


\end{document}